\documentclass[conference]{IEEEtran}
\IEEEoverridecommandlockouts
\usepackage{cite}
\usepackage{amsmath,amssymb,amsfonts}
\usepackage{algorithmic}
\usepackage{graphicx}
\usepackage{textcomp}
\usepackage{xcolor}
\usepackage{comment}
\usepackage{mathtools}
\usepackage{xspace}
\usepackage{comment} 
\usepackage{float} 
\usepackage{tikz}

\newcommand{\as}[1]{\textcolor{black}{{#1}}} 
\newcommand{\ass}[1]{\textcolor{red}{{#1}}} 

\newcommand{\al}[1]{\textcolor{brown}{{\bf #1}}} 

\def\BibTeX{{\rm B\kern-.05em{\sc i\kern-.025em b}\kern-.08em
\kern-.1667em\lower.7ex\hbox{E}\kern-.125emX}}

\usepackage{comment} 
\usepackage[english,ruled,linesnumbered,noline]{algorithm2e}
\usepackage{setspace}
\usepackage{caption}
\usepackage{subcaption}
\usepackage{array}
\usepackage{booktabs} 
\usepackage{bigstrut} 
\usepackage{boldline} 
\usepackage{multirow} 
\usepackage{threeparttable}
\usepackage[normalem]{ulem} 
\usepackage{anyfontsize}
\usepackage{hyperref}

\definecolor{ao(english)}{rgb}{0.0, 0.5, 0.0}

\definecolor{color1}{RGB}{198, 90, 103}
\definecolor{color2}{RGB}{175, 42, 48}
\definecolor{color3}{RGB}{0,128,128}
\definecolor{color5}{RGB}{128,0,128}

\definecolor{color3}{RGB}{0,0,0}

\newcommand{\cp}[1]{\textcolor{blue}{#1}}

\newcommand{\revisao}[1]{{\color{black}{#1}}}

\title{

Relational Consensus-Based 
Cooperative Task Allocation Management for IIoT-Health Networks
}

\author{
\IEEEauthorblockN{{\bf Carlos Pedroso\IEEEauthorrefmark{1}}, {\bf Yan Uehara de Moraes\IEEEauthorrefmark{1}}, {\bf Michele Nogueira\IEEEauthorrefmark{1}\IEEEauthorrefmark{2}}, {\bf Aldri Santos\IEEEauthorrefmark{1}\IEEEauthorrefmark{2}}}

\IEEEauthorblockA{\IEEEauthorrefmark{2}Department of Computer Science - UFMG, Brazil  \\
\IEEEauthorrefmark{1}Department of Computer Science - UFPR, Brazil \\
Emails: \{capjunior, yumoraes, michele, aldri\}@inf.ufpr.br 
}}

\IEEEaftertitletext{\vspace{-1.0\baselineskip}}

\newcommand\copyrighttext{%
	\centering
	\footnotesize \textcopyright~ IFIP, (2021). This is the author's version of the work. It is posted here by permission of IFIP for your personal use. Not for redistribution.
	The definitive version was published in Proceedings of IFIP/IEEE International Symposium on Integrated Network Management 2021
	}
\newcommand\copyrightnotice{%
	\begin{tikzpicture}[remember picture,overlay]
	\node[anchor=south,yshift=8pt] at (current page.south) {\fbox{\parbox{\dimexpr\textwidth-\fboxsep-\fboxrule\relax}{\copyrighttext}}};
	\end{tikzpicture}%
}

\begin{document}
\maketitle
\copyrightnotice
\thispagestyle{plain} \pagestyle{plain} 

\begin{abstract}
IIoT services focused on industry-oriented services often require objects run more than one task. IIoT objects poses the challenge of distributing and managing task allocation among them. The fairness of task allocation brings flexible network reconfiguration and maximizes the tasks to be performed. Although existing approaches optimize and manage the dynamics of objects, not all them consider both co-relationship  between  tasks  and  object capabilities and the distributed allocation over the cluster service. This paper  introduces the ACADIA mechanism for task allocation in IIoT networks in order to distribute task among objects. It relies on relational consensus strategies to allocate tasks and similarity capabilities to determine which objects can play in accomplishing those tasks. Evaluation on NS-3 showed that ACADIA achieved 98\% of allocated tasks in an IIoT-Health considering all scenarios, average more than 95\% of clusters apt to performed tasks in a low response time, and achieved 50\% more effectiveness in task allocation compared to the literature solution CONTASKI. 
\end{abstract}



\SetKwProg{Procedure}{procedure}{}{{end procedure}}
\SetKwFunction{FWarmUp}{WarmUp}
\SetKwFunction{FBroadcastCap}{BroadcastCapabilities}
\SetKwFunction{FReceiveCap}{RecvCapabilities}
\SetKwFunction{FSimCalc}{SimilarityCalculation}
\SetKwFunction{FSelectLeader}{SelectLeader}

\SetKwFunction{FRecvLeaderRegister}{APRecvLeaderRegister}
\SetKwFunction{FSendTask}{APSendTask}
\SetKwFunction{FRecvTask}{LeaderRecvTask}

\section{Introduction}
The Internet of Things (IoT) is a heterogeneous network whose objects own characteristics like identity, physical attributes, computational and sensing capabilities~\cite{gubbi2013internet,botta2016integration}. For the vast majority of IoT objects,
it is important to reduce power consumption while communicating or making  certain tasks. Meanwhile, objects must evenly share their resources and cooperate to support better network performance~\cite{khalil2019new}. Thus, objects that run a set of functions can collaborate to allocate and realize different tasks~\cite{ghanbari2019resource}. 
In this context, Industrial Internet of Things (IIoT) has 
drawn greater attention~\cite{xu2018industry} since it focuses on connecting 
multiple objects with numerous capabilities within an industrial environment, 
enabling thus every device to work in a synchronized and organized manner to perform sensing and monitoring tasks. Health application use the advanced technologies of IIoT to
create interaction between patients and medical staff, hospital, and medical devices by creating a smart environment for the health domain~\cite{qin2020recent, Rodrigues2018}. Reaching fairness
performance in the distribution management 
of~\revisao{multiple} tasks between IIoT objects is challenging due to 
configurations required by IIoT networks~\cite{aazam2018iot}.
Inefficient allocations
of sensing tasks between IIoT objects causes problems in distribution of computational resources, 
bad environmental setting, damaging the data collection and availability~\cite{yuehong2016internet,hasan2020task}. In this way, we must preserve the efficiency of the allocation of tasks so that the objects are always able in emergencies to perform the reconfiguration of the environment and also deal with the entire volume of data generated. The wrong configuration of the environment makes it difficult to classify objects and makes the volume of data generated by them available since it
takes a long time to reach the application and directly impacts its interpretation. In smart hospitals with high demand for multiple emergency services, 
the dynamic configuration between the environments and services must be done in an optimized, fast, distributed, and flawless manner~\cite{rahmani2018exploiting}. Thus, intelligent management optimizes the resources of IIoT objects in the various tasks, by allowing objects to interleave execution according to the needs of the application and by preserving their resources~\cite{hasan2020task}.

Task allocation services have been extensively studied in WSN, usually treated as resource allocation for extending the network life~\cite{pilloni2017consensus},  in  IoT networks, that issue becomes essential for the arise of new applications~\cite{ghanbari2019resource}. Among the works that deal with the task allocation, many of them apply object virtualization in task groups~\cite{khalil2019new} or distributed consensus~\cite{pilloni2017consensus}. Object virtualization applies 
assign tasks according to the sensing competencies 
of each object and its performance capacity
in order to optimize the task execution to save resources. The distributed consensus applies to the equitable distribution of resources based on the interactions of each group of objects present in the network, this approach is applied in networks with a large number of participants~\cite{colistra2014task}.

However, these solutions disregard the similarity relationship between objects and the executing tasks. Besides that, they do not take into account
an allocation decision based on the traits of the environment where objects are inserted. Hence, IIoT demands dynamic, adaptable, and fair solutions to handle a range of objects and to provide transparent~configuration. Moreover, the solutions need to disseminate tasks among objects aware of their relationship with the environment and also the object capabilities~\cite{qin2020recent} for getting balanced management of 
the IIoT network resource. 
Though, it is  crucial for IIoT to provide mechanisms able to manage task allocation among objects by 
their relationships and capabilities for more robust and fairness management of available network resources.

This paper introduces ACADIA (\textit{Rel\textbf{A}tional \textbf{C}onsensus-B\textbf{A}se\textbf{D} Task Allocation for \textbf{I}IoT-He\textbf{A}lth}) a mechanism
for supporting sensing task allocation among objects into IIoT-Health networks.
ACADIA arranges IIoT objects into similarity-based clusters to address an effective distribution of sensing tasks between available objects. This work extends to the health domain our previous work of task allocation in~\cite{Pedroso2020ISCC}, as well as addressing the existence of multiple tasks.  
ACADIA employs collaborative relational consensus for better adaptation on the context, quick responses, and getting assertive decisions 
about the capabilities of objects to make specific sensing tasks. 
It also allows us to
allocate simultaneously multiple and distinct tasks among the clusters.  In analysis  against CONTASKI~\cite{Pedroso2020ISCC} on the NS-3 simulator,
ACADIA achieved 98\% of suitably allocated tasks in a given IIoT-Health domain, average more than 95\% of clusters apt to realize sensing tasks in a low response time and, achieved 50\% more effectiveness in task allocation compared to CONTASKI.

This paper is organized as follows: Section~\ref{sec:rel} discusses the related work. Section~\ref{sec:sys} defines the model and assumptions taken by ACADIA.
Section~\ref{sec:CONF}  describes the ACADIA components and their operation.
Section~\ref{sec:ana} shows
the evaluation methodology to analyze the performance, and 
the results obtained. Section~\ref{sec:con} presents conclusions.

\section{Related Work}
\label{sec:rel}
The demand for dynamic and distributed services based on the resources and sensing capabilities of IoT objects has been the focus of several works~\cite{khalil2019new},~\cite{pilloni2017consensus},~\cite{colistra2014task}, ~\cite{pilloni2011deployment}. 
Though,  most of them 
still face 
many issues by managing the allocation  
of the IoT resources, like co-relationship between tasks and object capabilities, fairness in multi-tasks distribution, and flexible network reconfiguration.
In~\cite{khalil2019new}, 
 an evolutionary algorithm based on heterogeneity recognition heuristic in IoT networks addresses to ensure greater stability and operational periods of tasks according to the current demands. 
The algorithm creates collaboration between the functions of objects by taking the task’s demands and virtual objects groups selected to realize 
tasks and also reduce the energy consumption. In this solution, however, only a few objects are capable of performing the tasks attributed to network, as well as only two types of tasks can be done, which limits its use on networks including nodes with diverse capacities. In~\cite{pilloni2017consensus}, virtual objects (VO) in an IoT smart health network realize the allocation of sensing tasks by a decentralized strategy, where VOs negotiate among them to reach a consensus on the resource allocation of the health devices. Despite it meets certain fairness in the task allocation, they employs only the same type of objects to perform the tasks, and hence ignoring the different sensing capacities, the varieties of interactions, and the impact of the network size.

In~\cite{Pedroso2020ISCC}, we proposed the mechanism called CONTASKI  for allocating task in IIoT that arranges the network into similarity-based groups to handle the division of tasks to be allocated. Although CONTASKI makes use of a distributed consensus strategy for decision making about the better task distribution for making a given service, we have ignored the simultaneous allocation of multiple distinct tasks, even being a condition expected in real networks. 
In~\cite{colistra2014task},  it is proposed a consensus-based heuristic approach to make decision on  fault tolerance task allocation in IoT. The approach  applies the concept of task groups and objects, so that in each task group, objects are chosen as virtual and vice-virtual. The model partly gets a flexibility on the network configuration, but it needs periodically to exchange \textit{hello} messages, being  computationally costly.  In addition, they ignore the resources capabilities of nodes, which directly influence on the distribution of tasks. 
In~\cite{kim2015efficient}, authors converted
the task allocation issue in an IoT environment into an integration problem with a minimal degree variant to narrow the task allocation, and thus applying a genetic algorithm to reduce its execution time.
Further, the objects only communicate each other by gateways services responsible for managing the  interaction. However, the gateways restricts the relationships between nodes and can cause communication bottleneck depending on the network size. In~\cite{pilloni2011deployment}, an algorithm decomposes sensing tasks in a sensor network into distributed ones, by taking the energy consumption of each task in order to get better resource allocation. They also apply a centralizing entity for distribution of roles, making it costly during the network reconfiguration. In~\cite{jin2011intelligent}, an algorithm for adaptive task mapping in sensors works jointly  with the task scheduling based on a genetic~algorithm to extend the network lifetime. Despite relating tasks and object capabilities, the centralization of distribution of tasks overloads the transmission channel and delays the message delivery,
compromising the synchronization of tasks execution.

\section{IIoT Health Environment}
\label{sec:sys}
This section presents the structure  of the IIoT health network, the manner how the objects communicate each other, and  the model of sensing tasks in which ACADIA can run to realize the allocation management. We assume a hospital setting with multiple
wings that can span over a number of floors and heterogeneous devices (IIoT objects)
capable of making different classes of sensing relative to the building's environment
and
patients' physiological signals.

\subsubsection{\textbf{Network model}}
An IIoT network composed by a set of objects denoted by $N = \{ob_{1}, ob_{2},...,ob_{n}\}$ in an area $(X_{x}, Y_{y})$. All objects own an unique network identifier $Id$ and differentiate each other by the sensing capabilities, represented by 
set $C =\{c_1, c_2, c_3,.,c_n \}$, processing power 
and memory. The objects are fixed in the setting and evenly distributed in the coverage area of the network. Also 
the objects do not suffer from energy restriction due to the existence of energy source in the network. 
Objects take roles as common or leader objects and
Access Points (AP).

\subsubsection{\textbf{Communication model}} Communication among the objects takes place over the wireless medium via a shared asynchronous channel, in which connections are reliable, and therefore the objects do not present communication failures. Also all objects 
exchange messages on the network layer. Further, the data sensed by objects can be accessed by the application layer regardless of the location, using protocols such as CoAP. 

\subsubsection{\textbf{Sensing task model}} Each task represents a demand for sensing
the ambient and/or patients' physiological signals
and it requires a set of sensing capabilities, whose size is variable, that depends on the setting ACADIA runs.
Those tasks take place in a programmed manner with predefined time duration. A task $T$ is~\as{a tuple} $\{ T_{id}, C,\tau, q\}$, so that  $T_{id}$ is the task unique $id$; $C$ means the set of sensing capabilities required to make the task, $\tau$ denotes the time needed to realize it and $q$ the~\textit{per-}cluster quorum to perform the task. The sensing activity takes into account two classes:
the infrastructural one senses the environment values such as temperature and light; the physiological one senses physiological signals like heartbeat and blood pressure~\revisao{as in~\cite{Rodrigues2018}}. The AP device keeps track of tasks' status 
pending, when the tasks are queued; dispatched, when they are accepted by some cluster, and completed.

\section{ACADIA}
\label{sec:CONF}

\begin{figure}[ht]
	\centering
	\includegraphics[width=0.90\linewidth]{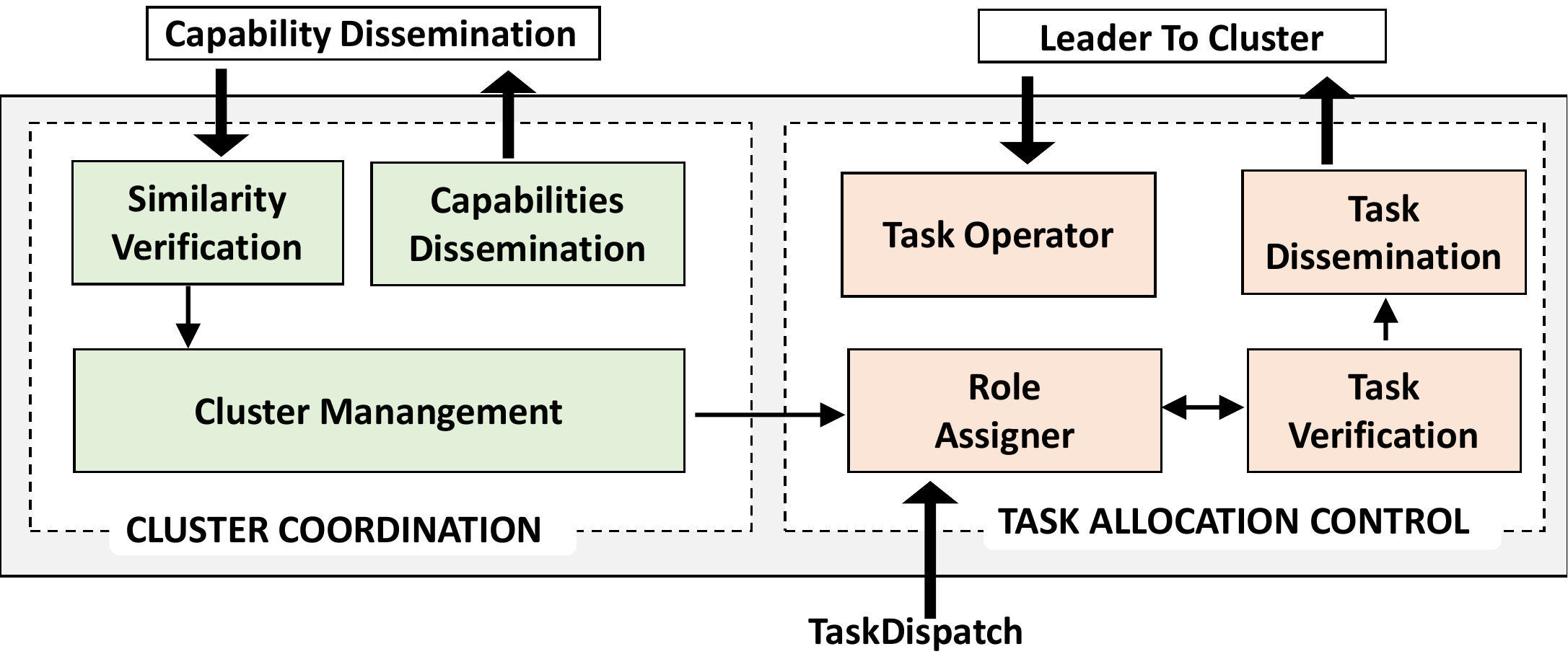}
	\caption{The ACADIA Architecture}
	\label{fig:ACADIA-architecture}
\end{figure}

The ACADIA architecture comprises two modules, called \textbf{Cluster Coordination} (\textbf{CC}) and \textbf{Task Allocation Control} (\textbf{TAC}), as shown  in Fig.~\ref{fig:ACADIA-architecture}. 
They act jointly to guarantee 
the configuration of clusters, as well as  
the dissemination and allocation of sensing tasks among IIoT-Health objects.
The \textbf{CC} module arranges the network in virtual clusters and the \textbf{TAC} module controls the dissemination of~\revisao{multiple} tasks to be done 
by the network objects
according to their sensing capabilities.
For achieving its goal, the modules exchange
five types of messages: \textit{CapabilityDissemination} 
that 
enable to configure the clustering; \textit{LeaderRegister} 
that make the leaders to register into AP; \textit{TaskDispatch} 
to dispatch tasks provided by the AP; 
\textit{TaskAccept} 
that support leaders to accept tasks; \textit{LeaderToCluster} 
that allow leaders to disseminate accepted task among the objects. 

The {\bf CC} module controls the creation and maintenance of the clusters by analyzing the neighboring objects using a similarity threshold of their capabilities in order to verify  if they are apt to participate in the same cluster. Therefore, 
every time that CC receives a~\textit{CapabilityDissemination} message come from the neighbors it verifies such data about its identification, capabilities and number of neighbors. 
For that, ~\textbf{CC} account with
three components:  
Capabilities Dissemination $(CD)$, responsible for disseminating~{\it CapabilityDissemination} messages with the object's  
identifier, its capabilities and number of neighbors; 
Similarity Verification $(SV)$, which receives and verifies the fields of messages exchanged among the objects; and  
Cluster Management $(CM)$, which manages the cluster creation using the objects' similarity, and 
the leader selection.

The {\bf TAC} module coordinates the sensing task allocation according to the object capabilities 
and dispatches
\revisao{multiple} tasks to the objects of the IIoT network\revisao{, deviating from~\cite{Pedroso2020ISCC},} in order to maximize and preserve their resources.
It comprises the following components: Task Verification $(TV)$, Role Assigner $(RA)$, Task Dissemination $(TD)$ and Task Operator $(TO)$. TV
oversees which tasks should be done and what capabilities are required by them. Whereas RA monitors the type of tasks to be assigned to the objects,
\revisao{employing}
relational 
consensus between the leader and
\revisao{tasks}
to evaluate which ones should be allocated according to 
\revisao{cluster's}
capability.
Next, DT dispatches the requested tasks considering 
the object 
capabilities.  
Lastly, TO takes care of the operation of 
the tasks
came from the leader.
Thus, the task allocation service becomes 
more fare and balanced, and does not overload the objects resources.

\subsection{Cluster configuration}
As the IIoT-Health 
size involves objects with different sensing
capabilities, the~\textbf{CC} module arranges the network objects in clusters based on leaders
in order to create a  network infrastructure capable to task allocate. 
Initially, objects begin cluster configuration exchanging \textit{CapabilityDissemination} messages that carries the $Id$, capabilities and number of neighbors of the sender. Algorithm~\ref{alg:cluster-configuration}
describes 
the cluster configuration process. Initially,  
each object sends a~\textit{CapabilityDissemination} message in order to announce its $Id$, sensing capabilities ($MyCapabilities$) and number of neighbors ($NeighborhoodSize$), through 
the procedure \FWarmUp.
When receiving a~\textit{CapabilityDissemination} message, the receiver updates its neighbors
($NeighList$), alongside with their capabilities ($NeighCapabilities$), by the procedure \FReceiveCap. The similarity verification takes into account those information, being calculated using cosine similarity. A neighbor can join into the cluster when its similarity is within the threshold. This update procedure occurs dynamically in all objects, ensuring that each one maintains its neighbor and cluster updated (procedure \FSimCalc).

\begin{algorithm}[hbt]
\relsize{-1.9}

\Procedure{\FWarmUp}{
    \While{60 seconds has not elapsed}{
	    \textsc{BroadcastCapabilities()} \\
	    \textsc{ReceiveCapabilities()} \\
	}
}

\Procedure{\FBroadcastCap}{
    $Broadcast(MyId, MyCapacities, NeighborhoodSize)$ \\
    $WaitInterval()$ \\
}

\Procedure{\FReceiveCap}{
    $Id \leftarrow GetId()$ \\
    $NeightList \leftarrow NeighList \cup Id$ \\
    $NeighCapabilities[Id] \leftarrow GetCapabilities()$ \\
    $NeighSize[Id] \leftarrow GetNeighborhoodSize()$ \\
}

\Procedure{\FSimCalc}{
    \ForEach{neighbor in NeighList}{
    	$NeighborCapabilities \leftarrow NeighCapabilities[neighbor]$ \\
    	$sim = \frac{|MyCapacities \cap NeighborCapabilities|}{\sqrt{|MyCapacities|*|NeighborCapabilities|}}$ \\
    	\If{$sim \geq Treshold$}{
    	    $cluster \leftarrow cluster \cup neighbor$
    	}
    }
}

\Procedure{\FSelectLeader}{
    $LeaderCandidates \leftarrow GetNeighborsGreatestNeighborhoodSize(cluster)$ \\
    $Leader \leftarrow GetNeighborLargestCapabilitiesSet(LeaderCandidates)$ \\
    \If{$Leader == MyId$}{
    	$SendLeaderRegisterToAP(MyId)$
    }
}

\caption{Cluster configuration}
\label{alg:cluster-configuration}
\end{algorithm}

The cluster leader selection (procedure \FSelectLeader)
takes into account~both the~number of neighbors and individual capabilities to choose the leader.
After that, the leader selected by the cluster informs to the AP that registers it as leader to guarantee the communication between AP,
leaders and cluster members and a better hierarchical network organization.
%

Equation~\ref{eq:similarity}
computes the similarity value
between two objects' capabilities, being based 
on~\cite[Eq.~3]{abderrahim2017ctms}.
The similarity takes into account the object's own capabilities ($C_{ob1}$) and the neighbor's capabilities ($C_{ob2}$). In this division, the upper part calculates
the norm of the vector that means 
the intersection between the capabilities. The bottom part takes the square root of the multiplication of the norm of each capability vector.
\begin{equation}
\label{eq:similarity}
sim(ob_1,ob_2) = \frac{|C_{ob1} \cap C_{ob2}|}{ \sqrt{|C_{ob1}|*|C_{ob2}|} }
\end{equation}

The similarity value varies from $0$ to $1$, being that closer to~$1$, more similar two objects are,  
being that, the similarity level is labeled 
\revisao{$S_1$ = Dissimilar, $S_2$ = Neutral and $S_3$ = Similar}
as a manner to show the levels of similarity objects and tasks get. This scale changes according to the previously established capabilities before the IIoT is deployed and modifies according to the demand of application.

\subsection{Task Allocation}
\revisao{Tasks are made available to carry out through the AP}, which keeps a list of
pending tasks and dispatches them according to 
settings' demands.
The AP sends group leaders the tasks via $TaskDispatch$ messages.
\textbf{TAC}
plays on an IIoT-Health infrastructure established
by the cluster configuration, and it runs guaranteeing resource maximization, i.e., allowing the task dissemination according to the capabilities of each cluster.
Algorithm~\ref{alg:task-allocation} describes the task allocation process and how leaders and the AP negotiate the tasks being performed.
In order to identify the leaders, the AP monitors the~\textit{LeaderRegister} messages and keeps its leader list updated~(procedure \FRecvLeaderRegister).
Initially,
the
AP manages
a collection of
pending tasks ($TaskList$).
When dispatching~\revisao{tasks}, the AP selects~\revisao{an amount} ($sm$) of multiple pending tasks from the list and sends a~\textit{TaskDispatch} message to the cluster leaders announcing the task $T$ to be executed~(procedure \FSendTask). After
each dispatch, it
waits
for a time interval for receiving the confirmation (\textit{TaskAccept} messages) from the available and compatible cluster leaders.
When  one confirmation is received at least, the AP
removes it
of the pending task list. 

Once the leaders receive the task $T = (T_{id},C,\tau, q)$, they verify the compatibility of 
their capabilities ($MyCapabilities$) 
with the capabilities $C$ needed to perform the task, and if the number of objects in the cluster is greater or equal the quorum $q$ needed.~\revisao{In case} they meet the criteria, the cluster leader confirms with the AP (\textit{TaskAccept} message) that it will perform the task and disseminates the task to the cluster. In case the cluster members cannot realize the task, the leader doesn't confirm this task with the AP~(procedure \FRecvTask).

\begin{algorithm}[ht]
\relsize{-1.9}

\Procedure{\FRecvLeaderRegister}{
    $LeaderId \leftarrow GetId()$ \\
    $LeaderList \leftarrow LeaderList \cup LeaderId$ \\
}

\Procedure{\FSendTask{sm}}{
    \ForEach{dispatch round}{
    	$DispatchedTasks$ = $DispatchMultipleTasks(sm)$ \\
    	\If{$WaitConfirmation()$}{
    		$TaskList \leftarrow TaskList - {DispatchedTasks}$
    	}
    }
}

\Procedure{\FRecvTask{$T = (T_{id},C,\tau, q)$}}{
    \If{
    	$C \subseteq MyCapabilities$
    	\textbf{and}  $|cluster| \geq q$}
    	{
    	    $SendTaskAccept(AP)$ \\
    	    $SendLeaderToCluster(T)$
    }
}

\caption{Task Allocation}
\label{alg:task-allocation}
\end{algorithm}

\subsection{Operation}
\begin{figure}[ht]
	\centering
	\includegraphics[width=0.95\linewidth]{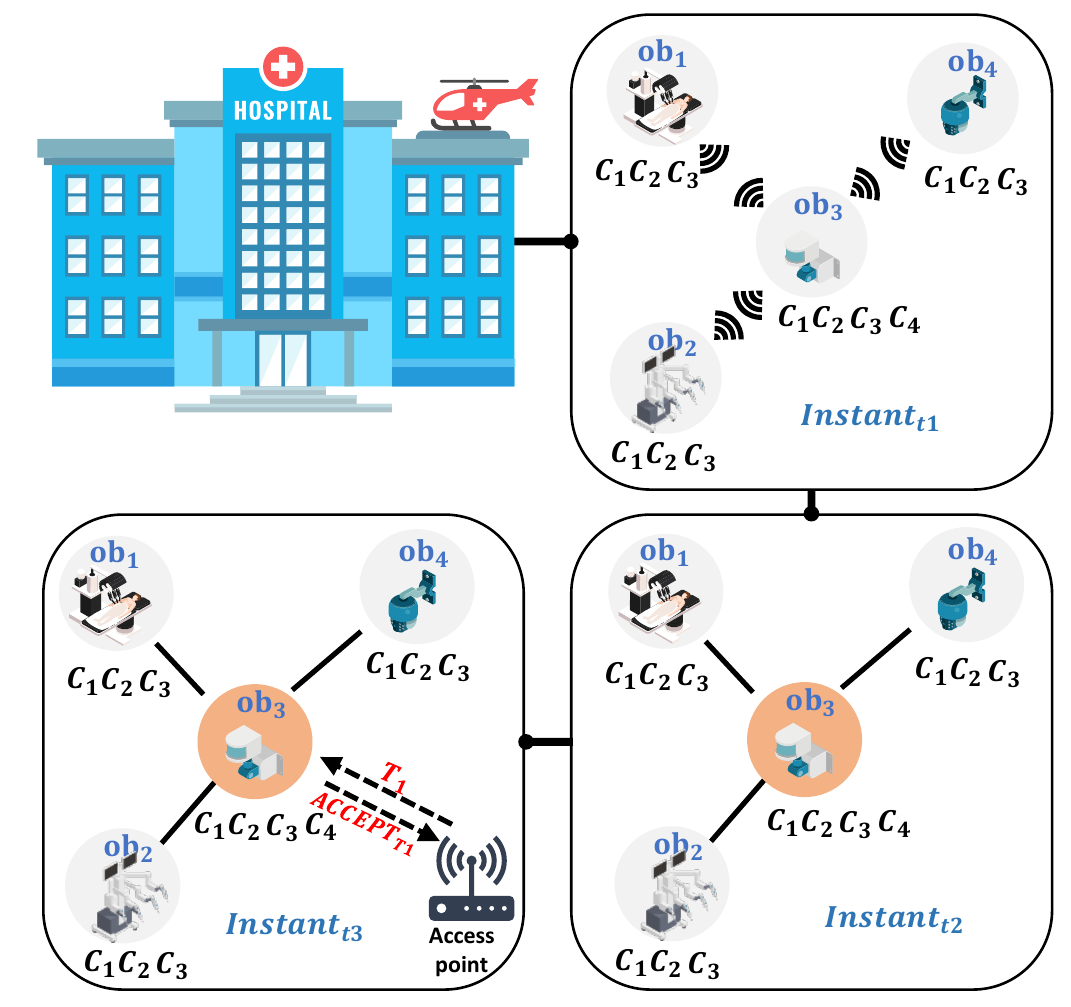}
	\caption{Formation of clusters and distribution of tasks}
	\label{fig:cluster}
\end{figure}

\revisao{ACADIA's task allocation acts dynamically and distributed in an IIoT-Health network on a hospital setting, 
whose objects are embedded in both medical equipment's and the structure of the building environment.}
The interactions between the IIoT objects occur over time and space dimensions, and objects in the transmission radius of the others exchange control messages in order to achieve a better configuration of the hospital activities.
Fig.~\ref{fig:cluster} illustrates how ACADIA acts for supporting the formation of IIoT objects cluster, leaders election, and allocation of tasks.The wireless signals mean objects within the transmission radius of each other and thus apt to exchange control messages about sensing capabilities. Each object carries its identifier $Id$ and a sensing capabilities set $C$ that it can make. Furthermore, capabilities $c_{1}$, $c_{2}$, $c_{3}$ and $c_{4}$ correspond to the sensing of temperature, humidity, lighting and body temperature\as{, respectively}. Besides, a similarity threshold ranging between $0$ (weak) and $1$ (strong) was setup for the formation of clusters, according to the capabilities of each object in the~network.

\as{We show the ACADIA operation on three different 
moments, said 
\textit{$I_{t1}$}, \textit{$I_{t2}$}, \textit{$I_{t3}$}} time instants. 
In~\textit{$I_{t1}$}, the set of~objects 
($ob_{1}$, $ob_{2}$, $ob_{3}$, $ob_{4}$)
exchange~\as{messages about its sensing capability and neighborhood} in order to realize the similarity calculation according to Eq.~\ref{eq:similarity}. The four objects obtain the following similarity values between them: 
$sim(ob_1,ob_2)=sim(ob_1,ob_4)=sim(ob_2, ob_4)=1$ and $sim(ob_1,ob_3)=sim(ob_2,ob_3)=sim(ob_3, ob_4)=0,87$.
As the similarities are within the range between $S_2$=Neutral and $S_3$=Similar,
objects in that interval
are clustered and
it means
objects with capabilities $c_{1}$, $c_{2}$,~$c_{3}$.

In~\textit{$I_{t2}$}, $ob_{3}$ is elected the cluster leader because 
it has higher number of neighbors than the others. With the cluster coordinator operating in this way, each object keeps its neighborhood information and capabilities updated through the exchange of messages.~Thus,
objects in the spatial neighborhood are seen as members of the same cluster, in addition to ensuring better scalability to the network, since a hierarchy based on leaders aids in the quality of the information transferred. Moreover, 
it facilitates the distribution of tasks among the objects of the network.
In~\textit{$I_{t3}$}, AP
dispatches a task to the leaders, that verify whether their cluster is apt to do it.
Over this view, $ob_{3}$, the leader, evaluates that the existing capabilities into the cluster are compatible and responds to the AP  confirming that the cluster will realize such 
task. 
Other clusters might exist in different points in the network that are capable of carrying out other tasks sent along with task $T_{1}$. 

\section{Analysis}
\label{sec:ana}
This section presents a performance evaluation of the ACADIA mechanism to assess its efficiency to the management of simultaneous tasks. We implemented ACADIA in NS3-simulator, version 3.29, and make all simulations taking into account an IIoT-Health scenario similar to a smart hospital with various levels. The object capabilities  are classified in structural and health sensing functions. The former follows the ones in~\cite{xu2019big} that consist of \textit{temperature}, \textit{humidity}, \textit{presence}, \textit{light}, \textit{position}, and~\textit{equipment condition}. The latter consists of \textit{heartbeat monitoring}, \textit{blood pressure}, \textit{body temperature}, \textit{oxygenation}, \textit{glucose level} and \textit{electrocardiogram} (ECG)~\cite{rahmani2018exploiting}. We evaluate six IIoT-Heath scenarios in the same aforementioned hospital setting, with 
50, 100 and 150 objects, and the number of simultaneous dispatched tasks, either 2 or 4 tasks, adding up to six scenarios configurations. 
The objects are evenly distributed in a area of~\textit{200 x 200} and communicate by IPv6 over IEEE 802.15.4. To avoid packet loss and bottleneck, we added a delay of~\textit{2ms} to messages exchanged among objects.

Scenarios with 2 dispatched tasks represent a configuration and sensing demand for an hospital wing with health services such as Emergency Response (ER), named ``\textbf{Demand A (\textit{D-A})}''.
In contrast,
scenarios with 4 dispatched tasks, the demand relates to comprehensive hospital services such as ER, Intensive Care Unit and Infirmary, named ``\textbf{Demand B (\textit{D-B})}''.
We randomly generate each task and their duration employing \textit{std::minstd\_rand0} as random generator. Tasks' total duration span~\textit{1380s} to account for multiple tasks being dispatched.
Seeds for each round constitute the sum of number of objects, run number and number of simultaneous dispatched tasks.
\textit{std::uniform\_int\_distribution} derive the objects capabilities and tasks using the random generator. The objects' capabilities, however, remained the same across the simulation rounds.
Randomly producing those values brings variation to the simulation to account for real-life differences. If tasks required the same capabilities, given that objects do not have mobility and their capabilities remained fixed across simulation rounds, then the error would be close to zero, and, therefore, not representative. But, this randomization also reflects on higher standard deviation, leading to high~amplitude~error~bars.

The system operates over {\it 900s} and in the first~\textit{60s} there is an exchange of messages between all objects to disseminate their capabilities, 
followed by the similarity computation and leader register. The AP dispatches multiple tasks every~\textit{60s}, from~\textit{60s} to up~\textit{840s}, and clusters perform them for~\textit{60s} or \textit{120s}, according to tasks' requirements. Moreover, each task has a random capabilities set, and all tasks require as structural capabilities at least \textit{temperature},~\textit{humidity} and~\textit{presence} and as health capabilities \textit{heartbeat monitoring},~\textit{blood pressure},~\textit{body temperature},~\textit{oxygenation}. 
The final capability set has up to three other structural capabilities and two other health capabilities, among the remaining ones of each type.

Pending tasks are forwarded to leaders by AP, always available, located in the center of the network, and equipped with a strong internet signal to reach all objects.
The similarity parameter varies from 0 to 1.
Clusters are classified as $apt$ and $inapt$ on each task dispatch, so that those $apt$ can make the dispatched task, and ones considered $inapt$ continue in an idle state saving resources for the next task. Also, considering the static scenario and transmission range of the objects, they form clusters with a non-deterministic number of participants. We also compare ACADIA  with the CONTASKI system~\cite{Pedroso2020ISCC} to analyze both performance in the task allocation. We assess the two systems with the following metrics based on~\cite{khalil2019new}:~\textbf{number of clusters (NC)},~\textbf{Number of unallocated tasks (NUT)},~\textbf{number of allocated tasks (NAT)},~\textbf{clusters apt to perform tasks (CPT)},~\textbf{clusters inapt to perform tasks (CIT)},~\textbf{latency of task accept time (LAT)} and,~\textbf{energy consumption (EC)}. 
All results 
correspond to the average of 35 simulations with a confidence interval of 95\%.

\subsection{Results} 

\begin{figure}[ht]
    \centering   
    \includegraphics[width=0.9\linewidth]{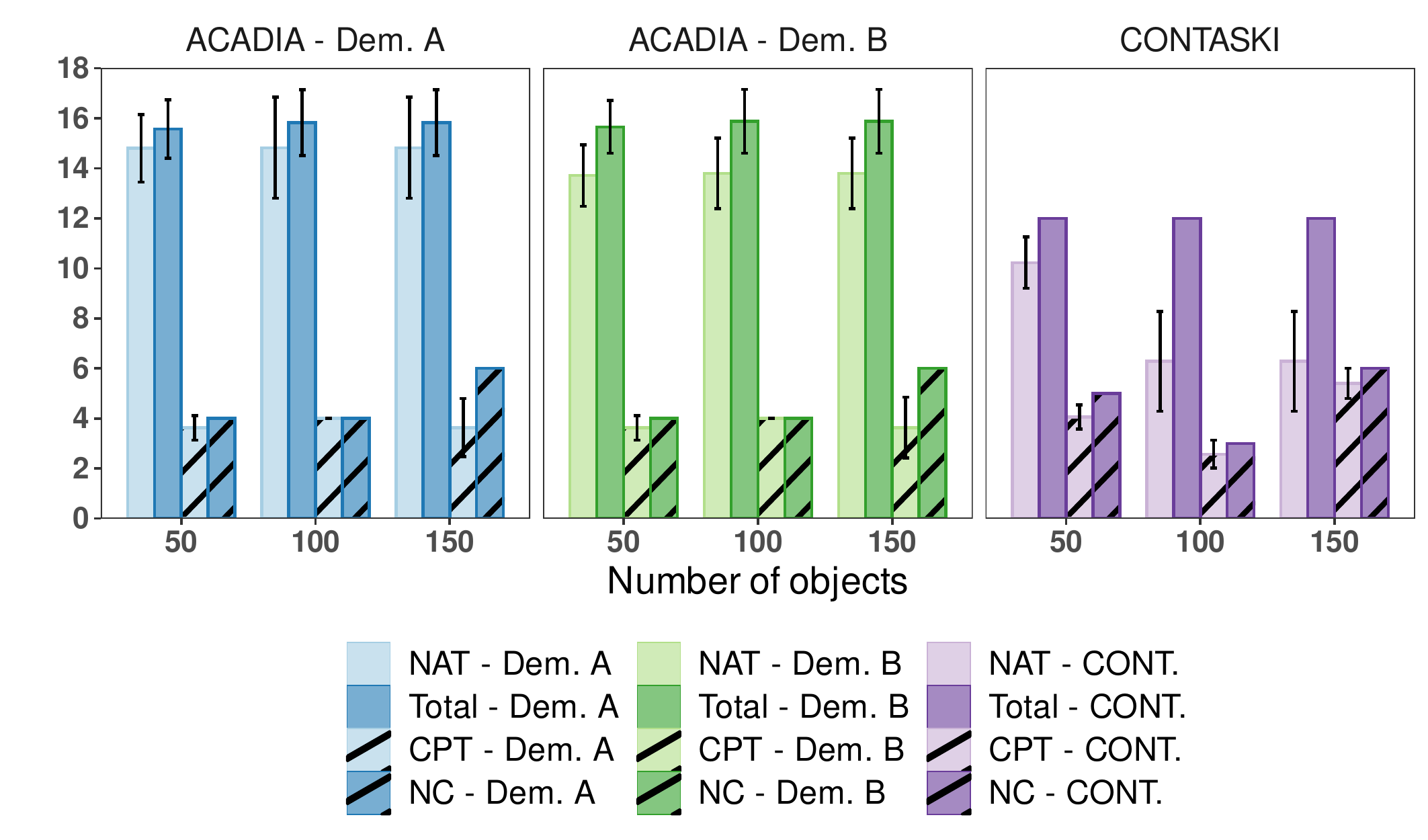}
    \caption{
        Apt Clusters (CPT) \& Task allocations (NAT)
    }
    \label{fig:nat-cpt}
\end{figure}

Fig.~\ref{fig:nat-cpt} (striped) shows  
the ACADIA performance for supporting the cluster formation, lighter striped bars represent the CPT and darker striped bars are the maximum. The CPT value relates to the similarity among objects that meets each capability set and the capabilities set of their neighbors, creating thus a consensual relationship between common objects and leaders to perform sensing tasks. ACADIA achieved an average CPT close to the average NC in most of the scenarios, showing that all clusters were apt to perform at least one of the dispatched tasks. 
In the scenario with 50 and 100 objects with either D-A and D-B
the NC remained close to 4, and the CPT closely follows that average. The 150 objects scenario showed an NC of 6 for both demands. However, 
the CPT value had an average of 3.8 clusters, reaching up to  
5 clusters. But, as it can be seen later, most of the tasks were performed. As for CONTASKI, it achieved higher CPT values, 4, 2.6 and 5.4, close to the averages of~5,~3 and~6. However, it did not translate in higher NAT values making explicit its inefficiency in task allocation.


Fig.~\ref{fig:nat-cpt} (non-striped)  exhibit  
the amount of sensing tasks dispatched by AP.
ACADIA's clusters were able to perform 98\% of dispatched tasks on average, with some rounds performing 100\% of them. But, ACADIA could not sustain 100\% of allocated tasks in all simulation rounds due to the random capabilities assigned to each object. Scenarios with {\it D-A} had over 98\% of allocated tasks and scenarios with {\it D-B} had over 86\% of allocated tasks.~In contrast,
CONTASKI allocated 100\% of tasks for the scenario with 50 objects and for others it only allocated 60\% of them, even though it had higher CPT values, revealing, again, inefficient task allocation. 
ACADIA was able to allocate around 100\% tasks, in all scenarios, obtaining 40\% more of NAT than CONTASKI. Moreover, it is noteworthy that CONTASKI dispatches only one task, whereas ACADIA dispatches 2 and 4 tasks simultaneously.

    
    
    

\begin{figure}[h!]
	\centering
	\begin{subfigure}{0.42\linewidth}
		\centering
	    \includegraphics[width=0.95\linewidth]{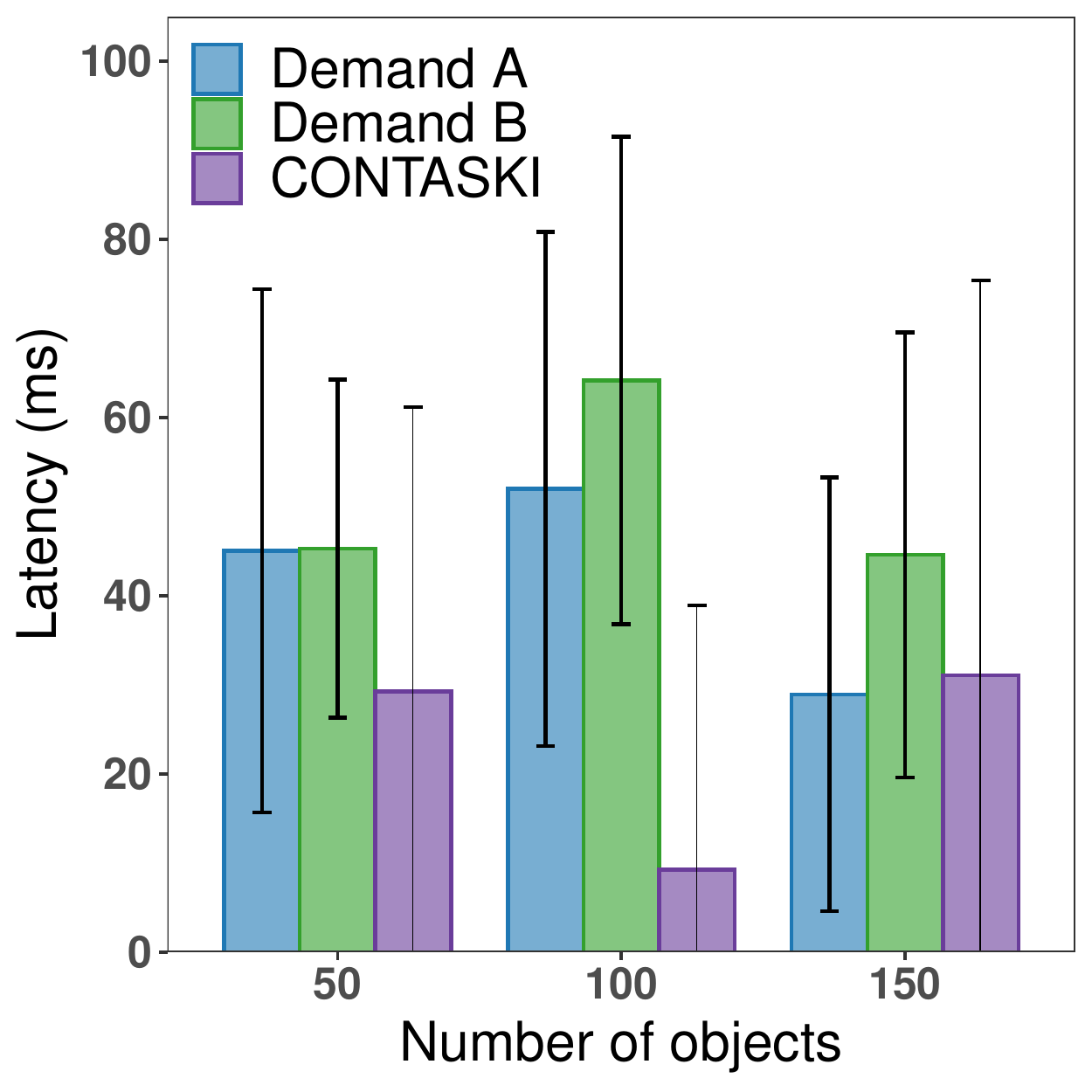}
		\caption{LAT}
		\label{fig:lat}
	\end{subfigure}
	\begin{subfigure}{0.42\linewidth}
		\centering
		\includegraphics[width=0.95\linewidth]{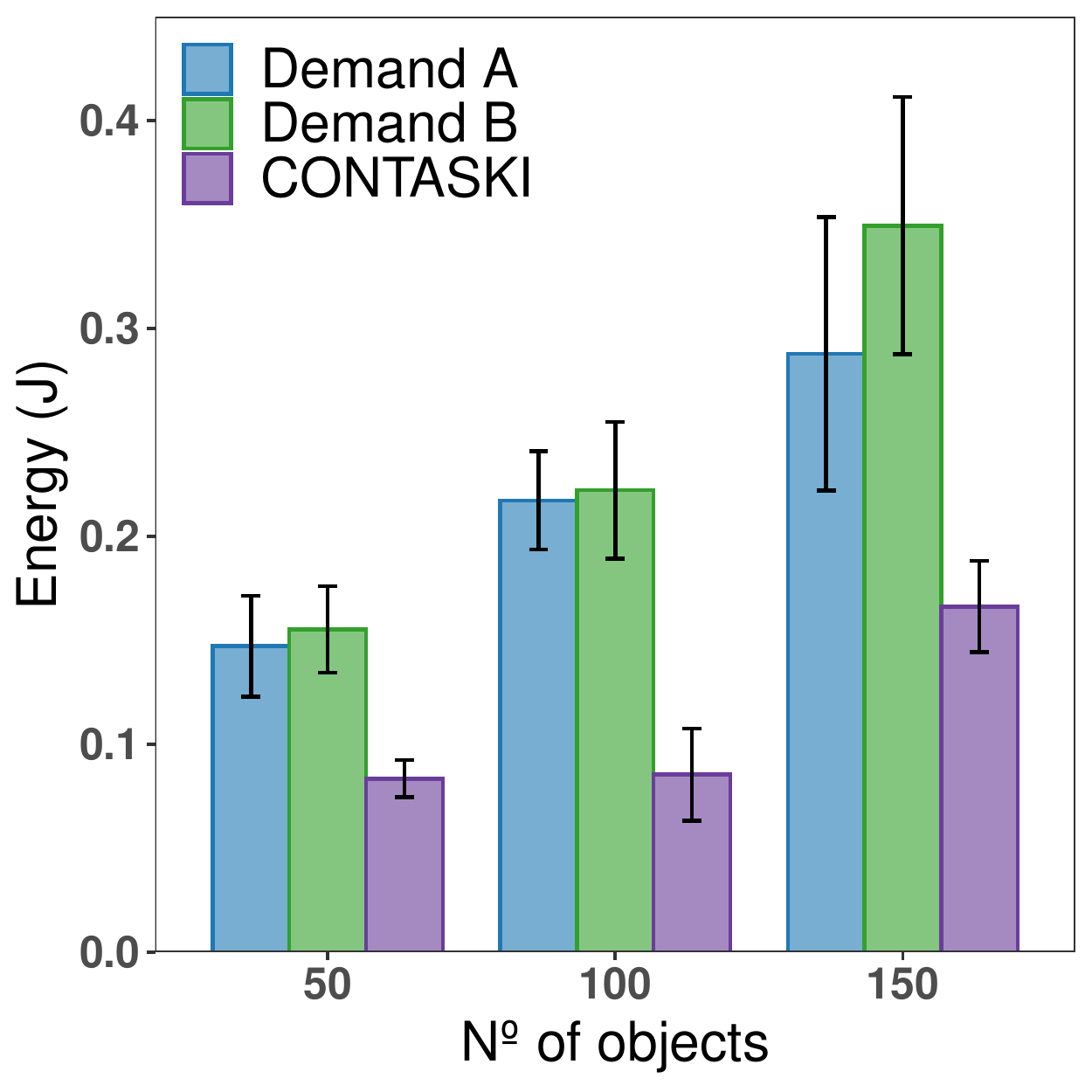}
		\caption{EC}
		\label{fig:acadia-energy}
	\end{subfigure}
\end{figure}

Fig.~\ref{fig:lat} shows the graphs of the latency of task accept time that quantifies the difference between the task dispatch time and the last accept time as seen by the AP.
ACADIA's \textit{D-A} and \textit{D-B}, with 50 objects, achieved similar LAT times around \textit{45ms}, given they had the same CPT.
The scenarios with 100 and 150 objects had the most variations between the demands due to factors mentioned previously. With 100 objects, \textit{D-A} achieved \textit{52ms} and \textit{D-B}, \textit{64ms}. In addition, with 150 objects LAT achieved \textit{28ms} and \textit{44ms}, respectively. Such variations observed are associated to the distance between objects and the AP, as well as the time taken by leaders to check it out if their capabilities are compatible to make the demand.
Further, CONTASKI shows poor performance in LAT times, since its standard deviation is so high, the error bars reach values below zero (not showed in the graph). ACADIA, in its turn, achieved consistent LAT times in all scenarios. CONTASKI's LAT times are \textit{30ms}, \textit{10ms}, and \textit{31ms}. ACADIA's higher LAT times -- in the order of 50\% higher than CONTASKI with 50 objects to almost four times with 100 objects -- however, translated in effective task allocation, as showed in NAT analysis.

    


Fig.~\ref{fig:acadia-energy} shows the graphs about 
ACADIA's and CONTASKI's energy consumption for the task allocation management, only including energy spent on message exchange.
It is noteworthy that ACADIA remained stable in energy consumption with little variation between the demands.
ACADIA spent~\textit{0.15}, \textit{0.22} and \textit{0.29} J in scenarios with 50, 100 and 150 objects in \textit{D-A}. The \textit{D-B} spent \textit{0.16}, \textit{0.22} and \textit{0.35} J, exhibiting less than 25\% variation between the demands.
While ACADIA exhibits a predictable trend in EC, CONTASKI 
displayed less energy consumption in all 
rounds. It spent \textit{0.08}, \textit{0.08} and \textit{0.02} J  exhibiting 80\%~\revisao{less} consumption then ACADIA's. However ACADIA's higher consumption translated close to 100\% of tasks allocated in most rounds.

\vspace{-0.2cm}
\section{Conclusion}
\label{sec:con}
This paper presented ACADIA for multiple tasks allocation on objects in an IIoT-Health network. It organizes the IIoT network 
into clusters based on the similarity of the capabilities of the objects and
the neighboring objects. The mechanism applies the relational consensus to manage and distribute tasks between the clusters, considering the capabilities they inform. 
Results show the effectiveness of ACADIA in task allocation among IIoT objects.
ACADIA was also compared to another task allocation mechanism, 
showing its effectiveness in multiple task allocation.
As future work, we intend to evaluate scenarios with different types of mobility, priority
in the task execution by the clustering and security concerns.



\section*{Acknowledgment}
We would like to acknowledge the support of the Brazilian Agency CNPq,  
grants \#436649/2018-7 and \#426701/2018-6.

\bibliographystyle{IEEEtran}
\bibliography{sbc-template.bib}

\end{document}